\documentclass[usenatbib]{mnras}
\usepackage{aasmacros}
\usepackage{amsmath,times}
\usepackage{graphicx,url}
\usepackage{amssymb}
\usepackage[usenames,dvipsnames]{color}
\usepackage{multirow}
\usepackage{natbib}
\usepackage{siunitx}
\usepackage{booktabs}
\usepackage[normalem]{ulem}
\usepackage{float}

\newcommand{\eagle}{\textsc{Eagle}}
\newcommand{\apostle}{\textsc{Apostle}}

\newcommand{\nexus}{\textsc{Nexus}}
\newcommand{\subfind}{\textsc{Subfind}}
\newcommand{\msun}{M$_{\odot}$}

\newcommand{\mvir}{$M_{200}$}
\newcommand{\mfof}{$M_{\mathrm{FOF}}$}
\newcommand{\dv}{$\delta_V$}
\newcommand{\vmax}{$V_{\mathrm{max}}$}
\newcommand{\rmax}{$r_{\mathrm{max}}$}

\title[A cosmological simulation of a gravitational lens]{A high-resolution cosmological simulation of a strong gravitational lens}

\author[Richings et al.]{\parbox{0.9\textwidth}{Jack Richings$^{1,2}$, Carlos Frenk$^1$\thanks{e-mail: {\tt c.s.frenk@durham.ac.uk}}, Adrian Jenkins$^1$, Andrew Robertson$^1$ and Matthieu Schaller$^3$}
\vspace{0.3cm}
\\$^1$ Institute for Computational Cosmology, Department of Physics, University of Durham, South Road, Durham DH1 3LE, UK
\\$^2$ Institute for Particle Physics Phenomenology, Department of Physics, University of Durham, South Road, Durham DH1 3LE, UK
\\$^3$ Leiden Observatory, Leiden University, P.O. Box 9513, 2300 RA Leiden, The Netherlands
}

\begin{document}

    \pubyear{2019}

  \maketitle

  \label{firstpage}

\begin{abstract}
  We present a cosmological hydrodynamical simulation of a
  $10^{13}$~\msun{} galaxy group and its environment (out to 10
  times the virial radius) carried out using the \eagle{} model of
  galaxy formation. Exploiting a novel technique to increase the
  resolution of the dark matter calculation independently of that of
  the gas, the simulation resolves dark matter haloes and subhaloes of
  mass $5\times10^6$~\msun{}. It is therefore useful for studying the
  abundance and properties of the haloes and subhaloes targeted in
  strong lensing tests of the cold dark matter model. We estimate the
  halo and subhalo mass functions and discuss how they are affected
  both by the inclusion of baryons in the simulation and by the
  environment. We find that the halo and subhalo mass functions have
  lower amplitude in the hydrodynamical simulation than in its dark
  matter only counterpart. This reflects the reduced growth of haloes
  in the hydrodynamical simulation due to the early loss of gas by
  reionisation and galactic winds and, additionally, in the case of
  subhaloes, disruption by enhanced tidal effects within the host halo
  due to the presence of a massive central galaxy. The distribution of
  haloes is highly anisotropic reflecting the filamentary character of
  mass accretion onto the cluster. As a result, there is significant variation in the number of structures with viewing direction. The median number of structures
  near the centre of the halo, when viewed in projection, is reduced
  by a factor of two when baryons are included.
\end{abstract}


\section{Introduction}
\label{section:into}


Compelling evidence for the existence of non-baryonic dark matter
particles is provided by the temperature structure of the cosmic
microwave background radiation \citep{planck2016} and supported by
observations of gravitational lensing \citep[see][for a
review]{massey2010}.  Measurements of the cosmic large-scale structure
set constraints on the properties of the particles.
Thus, the observed large-scale distribution of galaxies rules out hot
dark matter, that is, particles with large primordial thermal
velocities, as the main form of dark matter
\citep{frenk1983,white1983,white1984}. On the other hand, the data are in
excellent agreement with the cold dark matter (CDM) model, in which
the particles have negligible primordial thermal velocities
\citep{davis1985,springel2005, rodriguez2016}. Warm dark matter (WDM)
models represent the current upper bound on the primordial velocity
distribution of the dark matter particle. Testing these models serves
to constrain the properties of dark matter in the early Universe and
also to guide searches for the fundamental particle nature of dark
matter.

The main distinguishing difference between the CDM and WDM models is
the predicted abundance of structures on the scale of dwarf galaxies
and below \citep{colin2000, bode2001, lovell2012, schneider2012, kennedy2014}. Current WDM
models of interest, for example a 7 keV sterile neutrino\footnote{Such
  models are motivated by the observation of a
  3.5~keV emission line in the X-ray
  spectra of galaxies and clusters \citep{bulbul2014, boyarsky2014}.}, predict an exponential reduction
in the abundance of structure below a mass of approximately
$10^8$~\msun{} \citep{lovell2012,schneider2013,hellwing2016, bose2017,
  lovell2017}; by contrast, 
in the CDM model the halo mass function continues to increases towards
low masses \citep{diemand2007, springel2008}. Precise measurements of
the abundance of such low mass haloes would constrain WDM models and,
if they were shown to be absent, would conclusively rule out the CDM
model.

Galaxies cannot form in halos of mass $\lesssim 10^8$\msun{}
\citep{sawala2013,sawala2016,benitez-llambay2020} so these can only be
detected through gravitational lensing effects, particularly the
distortions they cause to the images of strong lensing arcs produced
by much more massive lenses such as groups and clusters of galaxies
\citep{koopmans2005}.  This method has already been used successfully
to detect a $1.9\pm0.1\times10^8$~\msun{} dark satellite and the
detection sensitivity is expected to reach $\sim 2\times 10^7$~\msun{}
\citep{vegetti2012}\footnote{The definition of mass in these papers
  assumes a truncated pseudo-Jaffe model and differs from the
  definition used in more recent gravitational lensing studies which
  is based on the NFW model.}. \citet{li2017} estimate that analysis of
about 100 strong lensing systems could conclusively distinguish CDM
from the 7~keV sterile neutrino WDM models, while samples of quadruply imaged quasars have already been used to infer that the halo mass function continues down to masses $\lesssim 10^{7.8}$~\msun{} \citep{2020MNRAS.491.6077G}.

\citet{li2017} and \citet{despali2018} based their predictions of the
subhalo and field halo contributions to the lensing signal on
dark-matter-only (DMO) simulations. It is now well established that
the inclusion of baryons in the simulations has important effects on
the population of small-mass subhalos orbiting in Milky Way mass
haloes \citep{donghia2010,sawala2017, garrison2017, richings2018},
leading to a reduction in the abundance of subhaloes near the centre
of the host of at least 50\%. The size of these effects in general
depends on the size and shape of the galaxy at the centre.  Haloes
that produce visible lens arcs are typically ten times more massive
than the Milky Way halo \citep{bolton2008} and the galaxies that form
at their centres are different in size and morphology to the Milky
Way.

Simulating $10^{13}$~\msun{} haloes with a small enough particle mass
to resolve the population of $10^7$~\msun{} subhaloes necessary for
strong lensing tests, whilst also including the effects of baryons at
sufficient resolution, is computationally prohibitive with conventional
techniques. Here we describe and implement a new technique for setting
up the initial conditions of a cosmological simulation, so that dark
matter particles outnumber gas particles by 7:1. This approach allows
us to resolve $10^7$~\msun{} substructures within a $10^{13}$~\msun{}
halo, whilst following the gas dynamics at the full resolution of the
high-resolution \eagle{} simulation, ${\sim}10^5$\msun{} \citep{schaye2015}.
In the simulation described here, the masses of dark matter and gas
particles are approximately equal.  This approach has the added
benefit of avoiding the spurious growth in the sizes of galaxies
described by \citet{ludlow2019}, caused by gravitational two-body
scattering of unequal-mass particles imparting velocity kicks to the
lighter particles.

This paper is arranged as follows: in \S\ref{section:simulations} we
describe the creation and testing of the initial conditions of our
simulation, as well as some key diagnostics of the completed
simulation. In \S\ref{section:halos} we examine the effect of both
baryons and environment on the abundance and properties of field
halos. This section also includes a discussion of the definition of
the mass of a halo. In \S\ref{section:subhalos} we study the abundance
and concentration of subhalos in the central halo of the
simulation. We also consider the variation in the observed abundance
of structure due to projection effects. We conclude in
\S\ref{section:conclusions}.

\section{Simulations}
\label{section:simulations}

The simulation was performed using the \eagle{} Reference model \citep{schaye2015, crain2015} with one exception: in addition to the fiducial star formation rate calculation, any gas particle reaching a density $n_\mathrm{H} > 10^4 \, \rm{cm}^{-3}$ was directly converted into a star particle..

\subsection{Candidate selection}
It is important that the halo and associated central galaxy selected
for resimulation be representative of those that produce observed
lenses. \citet{despali2017} identified a sample of halos in the
\eagle{} 100 Mpc simulation \citep{schaye2015} which have similar
properties to lenses detected in the SLACS Survey
\citep{bolton2006}. This was designed to detect bright,
early-type lens galaxies, the most suitable for detailed lensing and
photometric studies, at $z\sim 0.2$.

The following criteria were used: 
\begin{itemize}
    \item The halo is at a redshift of approximately $z=0.2$.
    \item The halo must be relaxed (according the criteria of \citealt{neto2007}).
    \item The halo has a virial mass between
      $10^{12}$--$10^{14.5}$~\msun{}. 
(Less massive halos will not produce visible Einstein rings.)
    \item The halo has a velocity dispersion of between
      160--400~km/s. 
      inside the half-mass radius\footnote{The half mass radius is
        calculated in projection, averaging over three orthogonal
        directions.}.  
    \item The central galaxy is an Elliptical. Specifically, at least
      25\% of all star particles inside 20 kpc must be
      counter rotating, where direction of rotation is given by the
      total angular momentum of all the star particles in this region.

\end{itemize}
From the sample of halos we select one object for resimulation. In the
\eagle{} 100 Mpc volume run with the \textsc{Reference} subgrid model,
the halo has a FOF ID of 129, a mass of $M_{200}=10^{13.1}$~\msun{},
and is located at at [89.742, 42.189, 94.507] Mpc. 

\subsection{Construction of initial conditions}
\label{section:zoom_ics}
We use a zoom simulation \citep{frenk1996} to study the selected
halo. This allows us to resolve the low-mass substructures relevant
for tests of the CDM model whilst minimising the computational
burden. We find all particles which are less than 5.5 Mpc from the
potential minimum of the halo at redshift $z=0.2$. 
We then identify these particles in the Eagle
simulation initial conditions and trace them back to their comoving coordinates at the Big Bang using the
Zel'dovich approximation \citep{zeldovich1970}. This defines the
region of space known as the Lagrangian region, which is the patch of
the universe from which our target halo will form.

To perform a zoom simulation, the Lagrangian region is populated with
particles which have smaller masses than the particles of the parent
simulation. The rest of the volume is populated with more massive
particles, present only to reproduce the correct large-scale tidal
forces without significantly increasing the overall computational
cost. The particles which populate the Lagrangian region must be
arranged such that {\em (i)}~the whole region has the mean density of
the universe, {\em (ii)}~ the configuration of particles is very close
to being gravitationally stable. Any instabilities in the initial
conditions which are not to due physical effects will lead to the
rapid growth of artificial structure.

For DMO zoom simulations, the Lagrangian region can simply be
populated with a uniformly-spaced grid. A common approach for
simulations using the smooth particle hydrodynamics (SPH) technique is
to take the uniform grid of DMO particles and split each particle into
a gas particle and a dark matter particle. The total mass of each pair
is kept the same as the DMO particle, and the particles are placed
such that their centre of mass is the same as the position of the DMO
particle.  In this setup there is one dark matter particle per gas
particle, and the ratio of the particle masses is determined by the
cosmological parameters of the simulation,
i.e. $m_\mathrm{DM}/m_\mathrm{gas}\equiv\Omega_\mathrm{DM}/\Omega_\mathrm{b}$. In the Planck 2015
cosmology, this means that each dark matter particle is 5.36 times
heavier than a gas particle.

\begin{figure}
    \includegraphics[width=\linewidth,trim=3.5cm 3.cm 3.5cm 0.4cm]{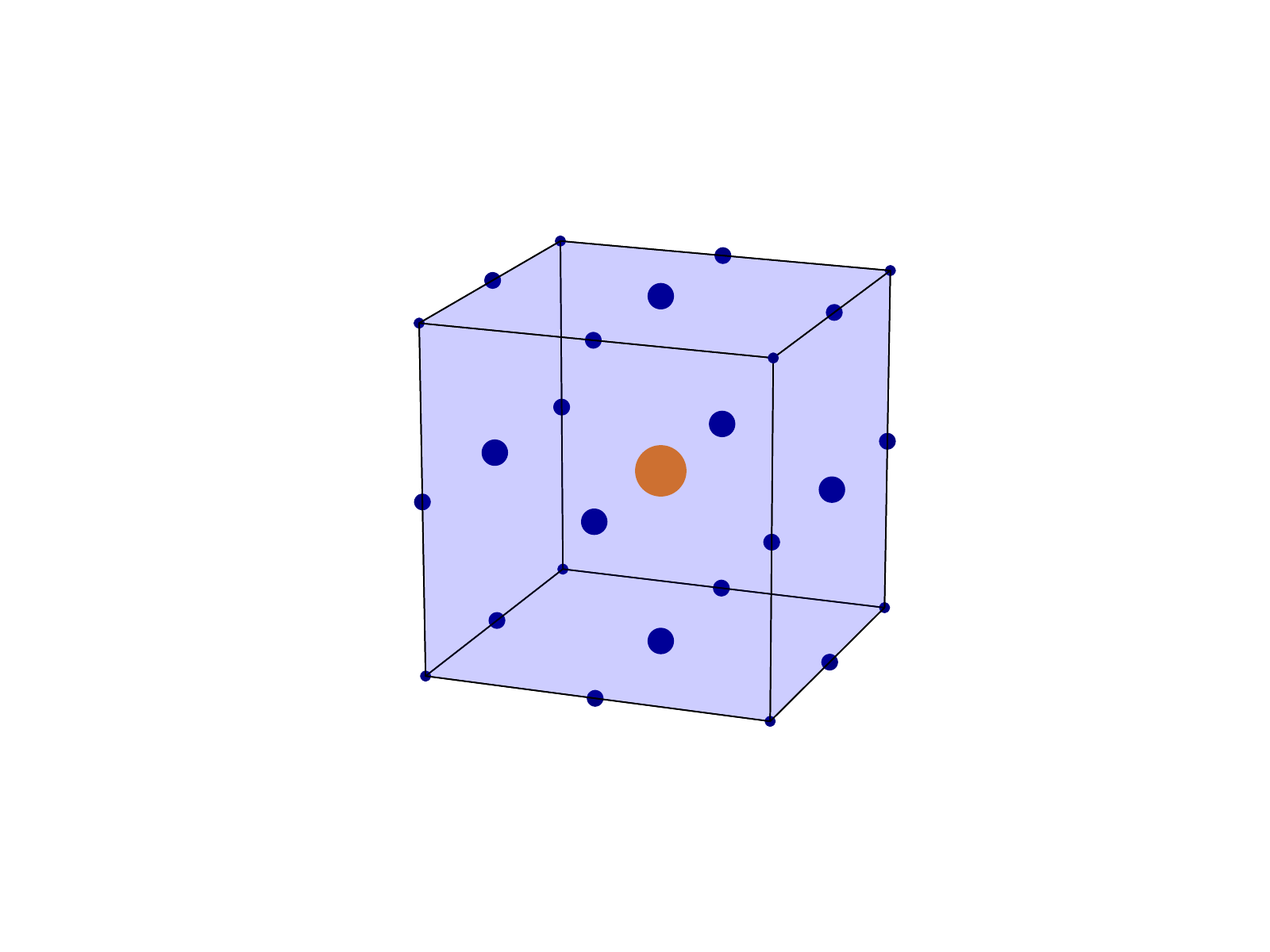}
     \caption{Template set of particles used to populate the Lagrangian
    region of the initial conditions. Dark matter particles are 
    blue, and the gas particle is orange. The area of each particle in 
    the diagram is directly proportional to its mass.}
    \label{fig:ic_cell}
\end{figure}

Our approach differs from the method outlined above in that the 
initial conditions are created with 7 dark matter particles per gas
particle. This means that the ratio of the particle masses is given by
$m_\mathrm{DM}/m_\mathrm{gas}\equiv\Omega_\mathrm{DM}/7\Omega_\mathrm{b}\sim0.77$. To ensure uniform
matter density, and to avoid gravitational instabilities (especially
at the boundary of the Lagrangian region), we tesselate the Lagrangian
region with a template as shown in Fig.~\ref{fig:ic_cell}. 

Each template contains one gas particle, which sits at the centre of
the cell. The template also contains 26 ``fractional'' dark matter
particles, positioned symmetrically on the faces, edges and vertices
of the cell. When two templates are placed next to each other, some
particles from each template will occupy the same position as
particles from the template next door. These coincident fractional
particles are combined into one whole particle, with a mass equal to
the combined mass of the original particles. In the interior of the
Lagrangian region, each face particle will overlap with one other face
particle, each edge particle will overlap with three other edge
particles, and each vertex particle will overlap with seven other
vertex particles. Therefore in order for the masses of all the dark
matter particles in the interior of the Lagrangian region to have the
same target mass, the mass of each face particle in the template is
one half of the target mass. Similarly, the edge and vertex particles
in the template have masses of one quarter and one eighth of the
target particle mass respectively. 

The total number of dark matter particles per template in the interior
of the Lagrangian region is thus given by $6/2 + 12/4 + 8/8 = 7$. Once
the Lagrangian region has been populated with copies of the template,
almost all dark matter particles will have the same mass, except for
dark matter particles at the boundary, which will have some fraction
of the target dark matter mass. These fractional masses at the edge of
the Lagrangian region are necessary to ensure uniform density and
gravitational stability. As the gas particle is placed at the centre
of the template all the gas particle masses in the Lagrangian region
will be the same.

Outside of the high resolution region, the tidal particles were placed using  the
method adopted for the Aquarius simulations \citep{springel2008}. Because the tiling
method for the high resolution is new, as a precaution,  we did an additional test on
the particle load. We created a full set of initial conditions with no cosmological
perturbations and ran a simulation from our intended start redshift 127 to redshift zero.

No structures formed within the high resolution region. Not
unexpectedly, some clustering occurred at the interface between the high
resolution region and the lightest mass tidal particles. This
structure formation, which is numerical in origin, was limited to a
thin surface only. The velocities remained small except close to this
surface. This indicates that the high resolution region in the
particle load is at precisely the mean density of the universe as
intended. The dark matter particles in this boundary region have
masses that differ from those in the interior of the Lagrangian
region. We excluded these particles, as well as all other tidal
particles from the analysis.
 
The cosmological parameters used for the simulation are taken from the
Planck 2013 results \citep{planck2014} and are listed in
Table~\ref{tab:cosmo}. The table also lists the gravitational
softening length used in the high-resolution region of our simulation
and the masses of the dark matter and gas particles in the initial
conditions.  The initial conditions contain about 198 million gas 
particles and 1.393 billion dark matter particles in the high
resolution region. In addition there are about 76 million more massive
`tidal' dark matter particles which surround the high resolution
region and fill the entire computational volume.
\begin{table}
\centering
\begin{tabular}{@{}ll@{}}
\toprule
Cosmological parameter                     & Value   \\ \midrule
$\Omega_m$                                 & 0.307   \\
$\Omega_{\Lambda}$                         & 0.693   \\
$\Omega_b$                                 & 0.04825 \\
$h\equiv H_0$/(100 km s$^{-1}$ Mpc$^{-1}$)  & 0.6777  \\
$\sigma_8$                                 & 0.8288  \\
$n_s$                                      & 0.9611  \\
$Y$                                        & 0.248   \\
$l_{\mathrm{box}}$ [cMpc]                   & 100   \\
$\epsilon_0$ [kpc]                           & 0.05   \\
$m_{\mathrm{DM}}$ {[}$10^4$ M$_{\odot}${]} & 8.27   \\
$m_{\mathrm{gas}}$ {[}$10^4$ M$_{\odot}${]}  & 10.74   \\ \bottomrule
\end{tabular}%
\caption{Cosmological and numerical parameters used in the simulation.
  $\Omega_m$, $\Omega_{\Lambda}$ and $\Omega_{b}$ are the mean density 
  of matter, dark energy and baryons in units of the critical density at
  redshift $z=0$; $H_0$ is the value of the Hubble parameter at redshift 
  $z=0$; $\sigma_8$ is the standard deviation of the linear matter 
  distribution smoothed with a top hat filter of radius 8 $h^{-1}$
  cMpc; $n_s$ is the index of the power law which describes the power 
  spectrum of primordial fluctuations; $Y$ is the primordial abundance 
  of helium; $\l_{\mathrm{box}}$ is the comoving side length of the 
  simulation box; $\epsilon_0$ is the softening length used in the force
  calculations for high-resolution dark matter and gas particles at redshift 
  $z=0$. $m_{\mathrm{DM}}$ is the mass of a dark matter particle in 
  the high-resolution region of the hydrodynamical version of the simulation.
  Edge effects in the construction of the initial conditions 
  mean that a tiny fraction of the high-resolution dark matter
  particles (approximately 1.5\%) have masses which are a fraction of this value. }
\label{tab:cosmo}
\end{table}
\subsection{Testing the initial conditions}

Changing the number of dark matter particles per gas particle can
potentially affect important observables in the final simulation. For
example, gravitational two-body scattering between species of
different masses influence observables like the size of small galaxies
\citep{ludlow2019}. To study the effects of increasing the dark matter
resolution for a fixed gas mass, we ran a 25~Mpc cosmological volume
with 376$^3$ gas particles and the same initial phases as the
L0025N0376 volume described in \citet{schaye2015}, but with seven
times as many dark matter particles. We refer to the original run as
the standard-resolution (SR) simulation, and our new volume as the
DMx7 simulation. The mass of gas particles in these two simulation are
the same, but our version has seven times as many dark matter
particles, that is our simulation has the standard \eagle{} gas
resolution, but a dark matter resolution similar to that of the
\eagle{} high-resolution (HR) run (L0025N0752).
\begin{figure}
  \includegraphics[width=\linewidth]{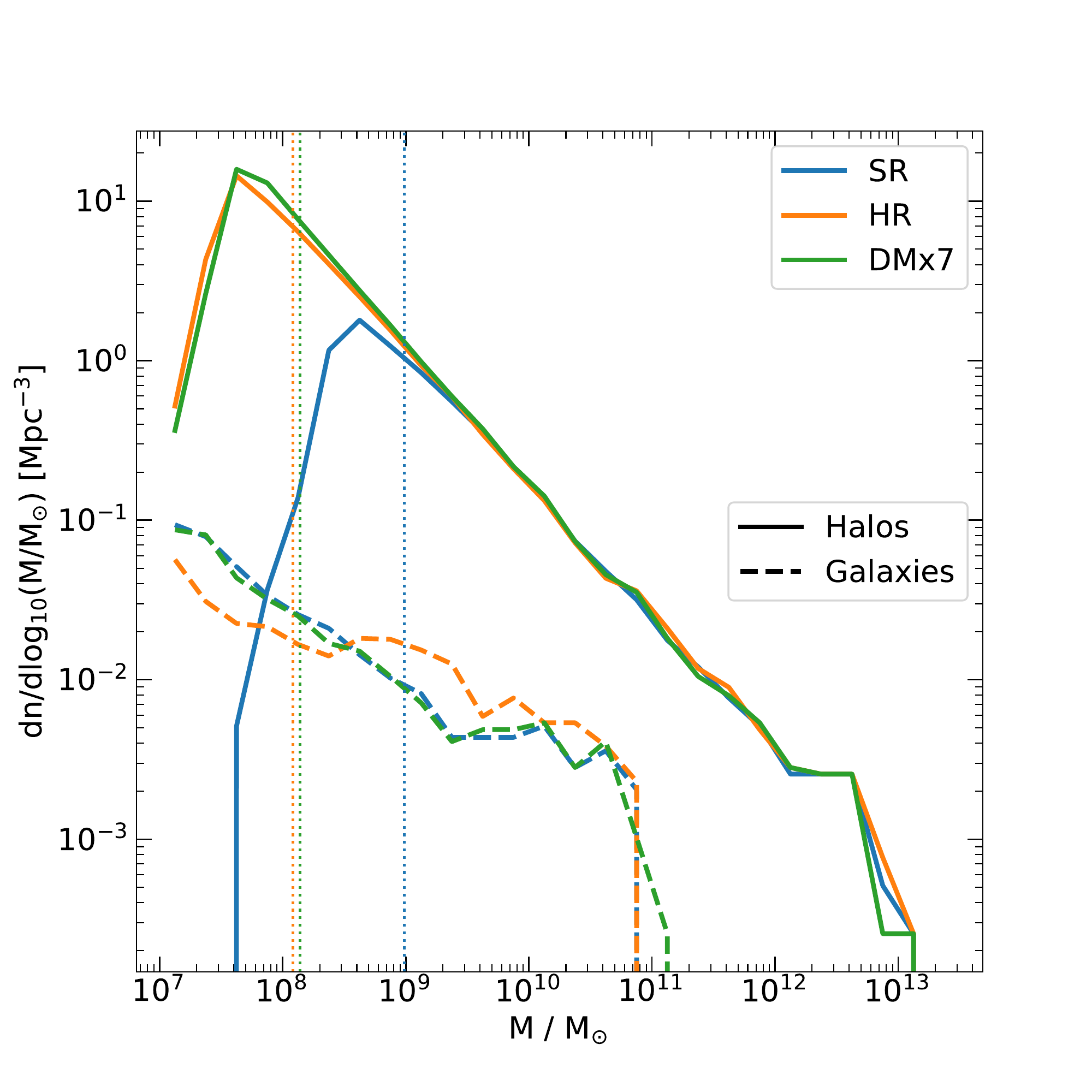}
  \caption{The mass function of halos (solid lines, $M=M_{200}$) and galaxies
    (dashed lines, $M=M_\star(<30 \, \mathrm{kpc})$) in three realisations of the \eagle{} 25~Mpc
    simulation. The blue lines show the halo and galaxy mass functions
    at standard \eagle{} resolution. The orange lines show the effect
    of increasing the resolution of both gas and dark matter in the
    simulation, while the green lines show the effect of only increasing
    the resolution of dark matter whilst holding the gas resolution
    constant, as described in \S\ref{section:zoom_ics}.  Dotted lines show the mass
    of 100 DM particles in each simulation.}
  \label{fig:testing_mf}
\end{figure}

We checked several key properties, the first of which is the mass
function of halos and galaxies. Here we take the mass of a galaxy to
be the mass of all star particles within 30 kpc of the potential
minimum of the host halo. These properties are shown in
Fig.~\ref{fig:testing_mf}. The mass function of galaxies is almost
unchanged between the versions of the simulation which have the same
number of gas particles but different numbers of dark matter
particles. The effect of increasing the resolution of gas particles
has a much more significant impact on the abundance of both smaller
and larger galaxies. The DMx7 simulation also does an excellent job of
reproducing the halo mass function at masses below the resolution
limit of the SR simulation. In general, if the difference between the
blue and orange lines is bigger than the difference between either
blue-green or orange-green, we conclude that the effect of increasing
the gas resolution is more significant than the effect of changing the dark
matter-gas mass ratio.
\begin{figure*}
  \includegraphics[width=\linewidth]{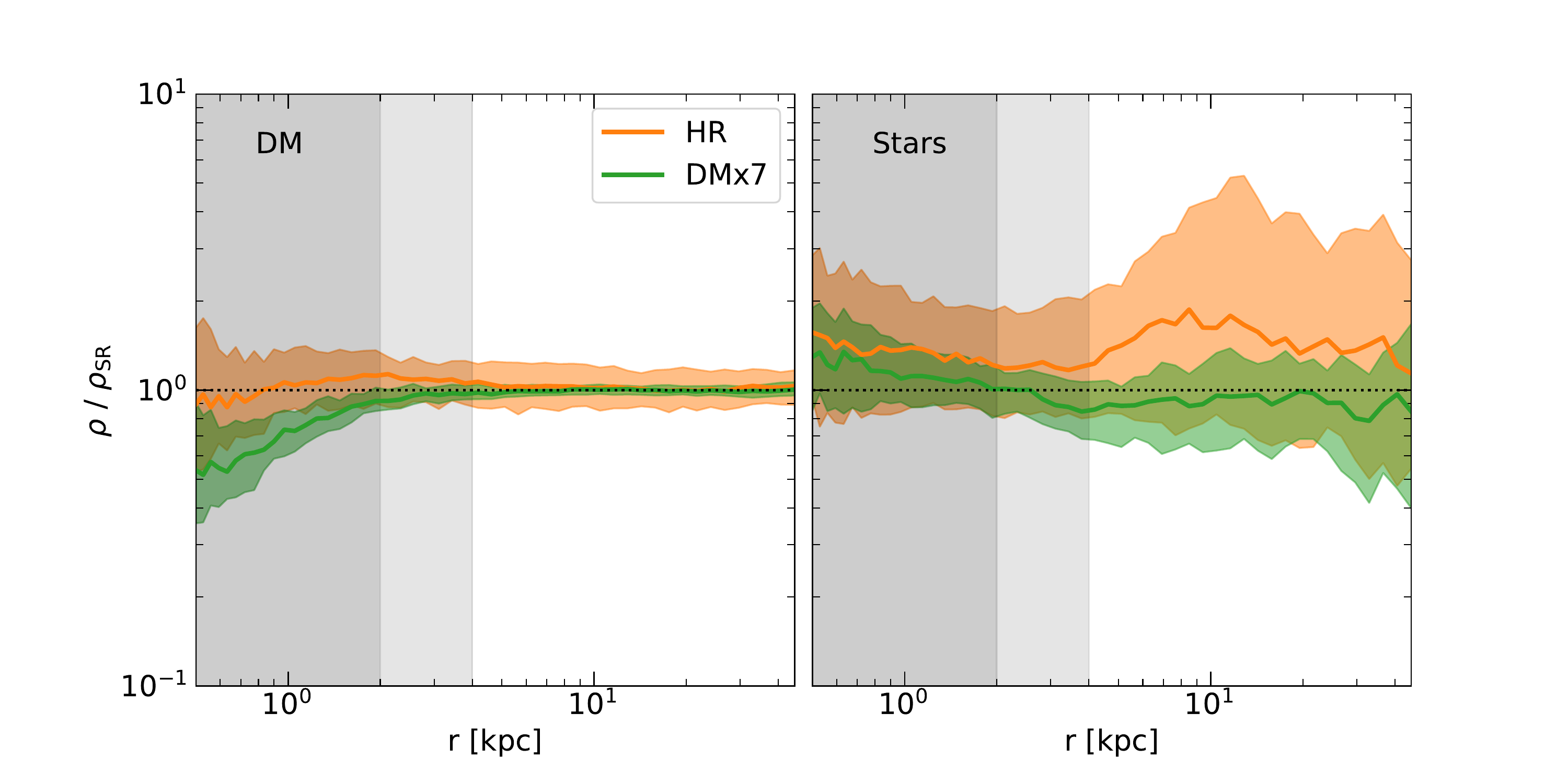}
  \caption{The ratio of the density of dark matter and stars in
    the HR (orange lines) and DMx7 (green lines) simulations to the
    density of dark matter and stars in the \eagle{} SR 25~Mpc
    simulation. The sample contains 100 halos bijectively matched
    between simulations. Solid lines show the median ratio as a
    function of radius for each species. Shaded regions indicate the
    interquartile range. The light grey shaded region shows the
    approximate value of the \citealt{power2003} radius for the SR
    simulation, whilst the dark grey region shows the corresponding
    radius for the HR and DMx7 simulations.}
  \label{fig:testing_den}
\end{figure*}

We also tested the effect of differing species resolution on the
internal structure of halos. We matched halos between simulations by mass and
position. Specifically, the masses of a potential matched pair must be
within a factor of two,\footnote{Typically the masses of matched halos agree to better than 10\%.} and the first halo must lie within the virial
radius of the second  halo and vice versa. This procedure produces a
unique match for each of the 100 most massive halos in the SR
simulation. Each halo in the SR simulation has a corresponding matched
halo in the HR and DMx7 simulations. We calculated the density of dark
matter and stars as a function of radius in each halo. For each
species, we then calculate the ratio of density in the HR and DMx7 to
the density in SR simulation. We performed this calculation for the 100
most massive halos in each simulation, which span a mass range
of approximately $10^{11.5}$--$10^{13.5}$~\msun{}. The results are
shown in Fig.~\ref{fig:testing_den}. 

Outside the \cite{power2003} radius, all three versions of the
simulations display excellent agreement in the measured dark matter
density profiles. At distances of less than 5 kpc from the centre of
the halo, the density of dark matter in the DMx7 simulation is
significantly lower than in the simulations which have a standard gas to
dark matter particle mass ratio. This result is not
unexpected. \citet{ludlow2019} have shown that the equipartition of
energy between multiple species of different-mass particles causes the
heavier species to sink artificially towards the centre of the
halo. In the case of the SR simulation, the dark matter particles are
around five times heavier than the star particles, which causes an
artificial increase in the density of dark matter at the centre of the
halo.

The second panel of Fig.~\ref{fig:testing_den} shows that beyond the
\cite{power2003} radius, where energy equipartition can affect the
distribution of particles, the density of stars is generally well
reproduced in the DMx7 simulation, albeit with considerable
scatter. The same cannot be said for the HR simulation, where the
effect of increasing gas resolution has a pronounced effect on the
distribution of stars in galaxies. The key takeaway is that
the uncertainties in the modelling of baryonic effects are
significantly larger than variations introduced by altering the ratio
of the mass of dark matter and gas particles.

\subsection{The simulation}
\begin{figure*}
\includegraphics[width=\linewidth]{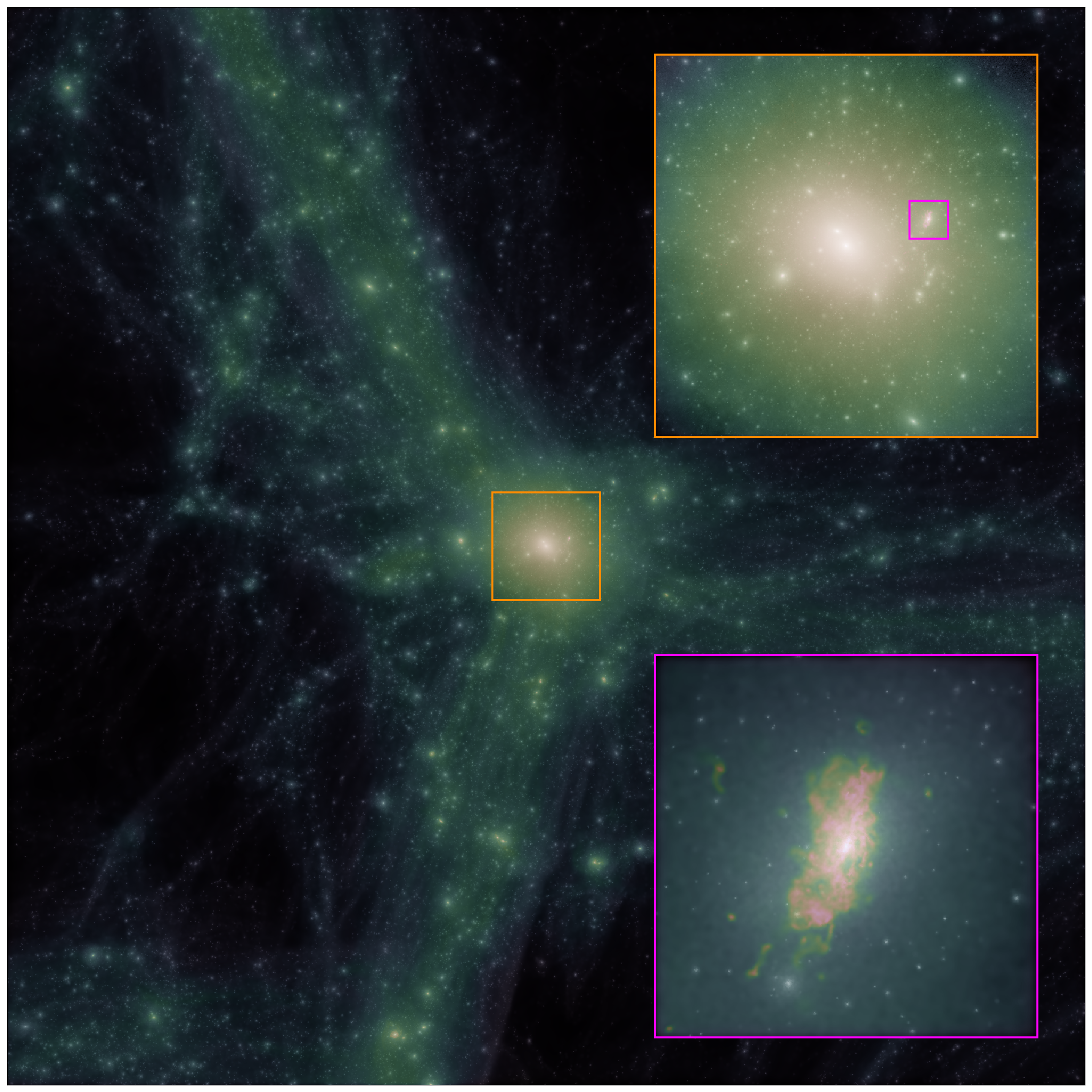}
  \caption{Projected density of matter in a cube of side length 10
    Mpc, centred on the most massive halo in the high-resolution
    region of the simulation. The brightness of each pixel is
    proportional to the logarithm of the density of matter, and the
    hue encodes the density of gas.  The orange inset shows a zoom
    into the largest halo, with a side length of 1 Mpc, and the pink
    inset shows a zoom into the subhalo with the greatest baryonic
    mass in the main halo, with a side length of 100 kpc.  The main
    image contains approximately 500 million particles, whilst the
    image in the pink inset is based on approximately 1.8 million
    particles.}
  \label{fig:simpic}
\end{figure*}
A visualisation of the high-resolution region of the simulation is
shown in Fig.~\ref{fig:simpic}. The brightness of each pixel in the
image is proportional to the logarithm of the projected density of
matter, in a cube of side length 10 Mpc. The projected density of gas
in the simulation is encoded in the hue of each
pixel. Fig.~\ref{fig:simpic} shows that the main halo in our
simulation sits at the centre of three large filaments. The inset
panels demonstrate the large dynamic range of
the simulation, with the volume of the cube shown in the pink square
being a millionth of the volume shown in the main figure. In addition
to the excellent resolution of the central halo, our simulation also
resolves the internal structure of the filaments of the cosmic web,
including strands of filaments that are almost entirely devoid of
baryonic matter. 

The region simulated at high resolution is unusually large for a zoom
simulation. The region is approximately spherical, with a radius of
around 7 Mpc at redshift $z=0$. This is approximately 14 times the
virial radius of the main halo. For comparison, the high-resolution
region in the Hydrangea cluster simulations is 10 times the virial radius \citep{2017MNRAS.470.4186B} and  is 4-5 times the virial radius 
(or around 1 Mpc in absolute terms) for the \textsc{Auriga} suite of galactic zoom simulations
\citep{grand2016}. The largest halo in the
high-resolution region (to which we will hereafter refer as the main
halo) has a mass of $M_{200}=10^{13.14}$~\msun{} and a radius of
$r_{200}=506$~kpc at redshift $z=0$. This halo contains 200 million
particles (as identified using the standard friends-of-friends
algorithm; \citealt{davis1985}). Running this hydrodynamical version of
this simulation required around 1.5 million core-hours, on 512 cores.

\section{The halo population}
\label{section:halos}

In this section we examine the field halos in our simulation. In
particular, we focus on the halo mass function in the mass range
$10^{6.5}$--$10^{10.5}$~\msun{}, critical for studies of strong
gravitational lensing by massive elliptical galaxies designed to test
the $\Lambda$CDM model and to distinguish CDM from viable alternatives
such as WDM in the form of 7~keV sterile neutrinos. 
We discuss the effects of baryons on the halo mass function, and
compare the measured halo mass function to predictions of the widely
used Sheth-Tormen model \citep{2002MNRAS.329...61S}. We also study the
relationship between halo properties and their environment,
specifically the abundance of halos in different environments and the
relationship between halo environment and internal halo structure. 

\subsection{The mass of a halo}

There is no unique way to define the mass of halos in cosmological
simulations. A number of definitions are widely used in the analysis
of simulations, and here we adopt $M_{200}$ -- the total mass contained
inside a sphere within which the mean density of matter is 200 times
the critical density of the universe -- as our definition. For each
halo, this sphere is centred on the particle in the corresponding
friends-of-friends (FOF) group \citep{davis1985} that has the lowest
gravitational potential. This means that there is one halo per FOF
group.

Several previous studies of the halo mass function have used the total
mass within each FOF group as the definition of halo mass
\citep{jenkins2001, springel2005, hellwing2016} and, when the FOF mass
is used, the Sheth-Tormen prediction for the halo mass function agrees
well with simulations.  Studies predicting the contribution to
strong-lensing perturbations from halos along the line-of-sight have
used the Sheth-Tormen mass function.  \citep{li2017,
  despali2018}. However, \citet{tinker2008} argue strongly in favour
of using a spherical overdensity method for measuring the mass of a
halo, as observable properties are more strongly correlated with
spherical overdensity masses than FOF masses.

We calculated both FOF and $M_{200}$ masses for the halos in the
high-resolution region of our simulation and found that $M_{200}$ is
typically lower than $M_\mathrm{FOF}$. For halo masses above
$10^9$\msun{}, the median ratio is approximately 0.9
\citep{jiang2014}, but at lower masses the discrepancy grows. This
implies that the mass function has a slightly shallower slope when
considering $M_{200}$ rather than $M_\mathrm{FOF}$, which leads to the
Sheth-Tormen mass function overpredicting the number of low $M_{200}$
mass halos.

\subsection{The effect of baryons on the halo mass function}
\label{section:baryon_mf}
A significant fraction of the distortions of strong lensing arcs is 
expected to come from halos along the line-of-sight, as opposed to
subhalos around the main lensing galaxy \citep{li2017}. It is
computationally difficult to simulate cosmological volumes on the
scale of hundreds of megaparsecs with sufficient resolution to
characterise the distribution of the low-mass halos of interest for
tests of the CDM model. As such, it is necessary to use an analytic 
prescription for the abundance of field halos when calculating the
expected lensing signal. 

Both \citet{li2017} and \citet{despali2018} used the analytic
Sheth-Tormen mass function to predict the number of halos lying
between the source galaxy and the observer (so-called
interlopers).\footnote{\citet{despali2018} used updated values for
  some of the numerical parameters related to the Sheth-Tormen mass
  function. These updated parameters provide a better match to the
  mass function in simulations with a \emph{Planck} cosmology
  \citep{2016MNRAS.456.2486D}}
%
\begin{figure}
  \includegraphics[width=\linewidth]{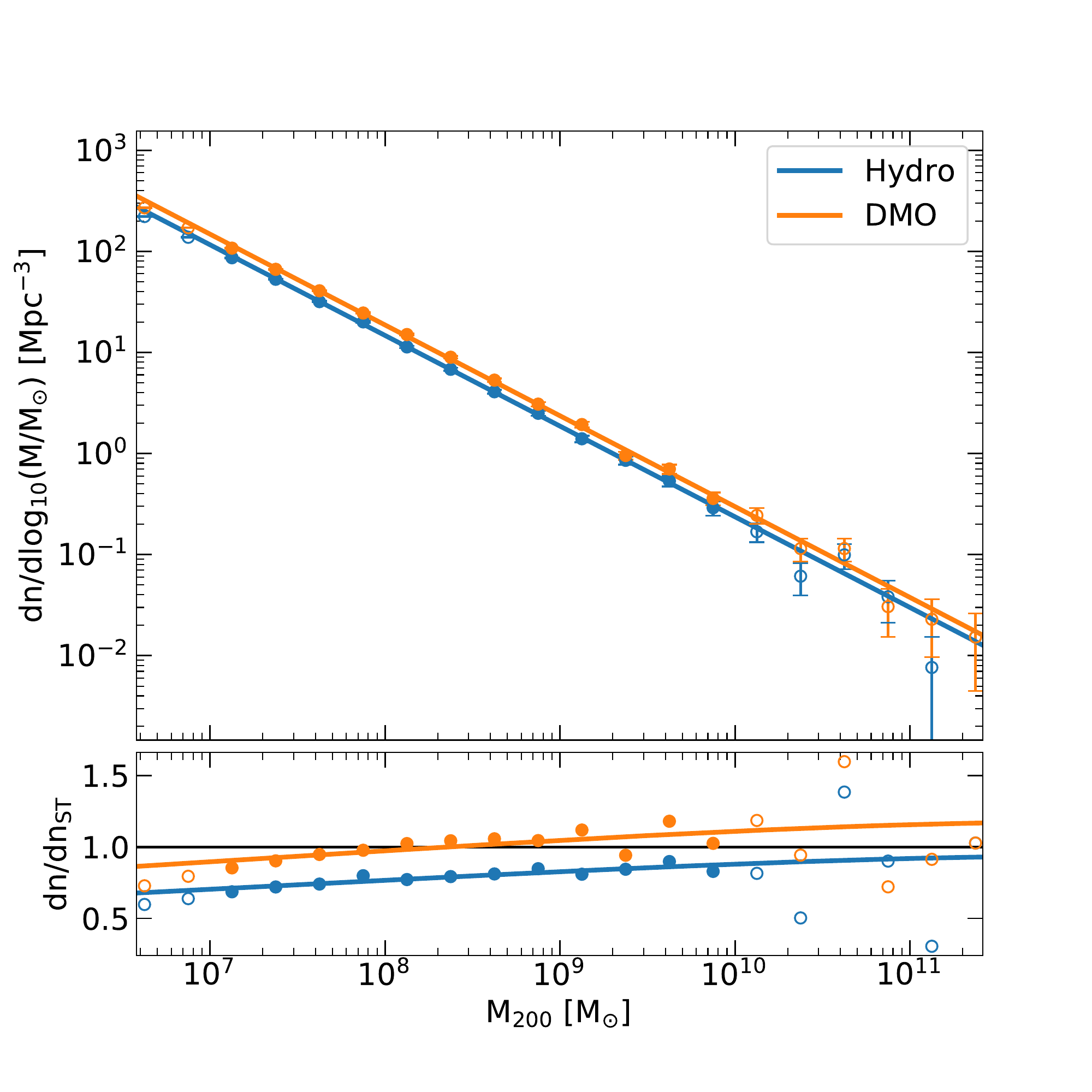}
    \caption{Top panel: the differential mass function of field halos in the hydrodynamical
    and DMO versions of our simulation, shown in blue and orange respectively.
    The mass function is calculated in a sphere of radius 5 Mpc centred on
    the potential minimum of the most massive halo in the high resolution 
    region of the simulation. Circles show the measured halo
    mass function in each mass bin. The errorbars show the Poisson
    error. Solid lines show power-law fits to the halo mass function.
    Points shown with empty circles were not used  when calculating the power-law fit.
    Bottom panel: the ratio of the calculated halo mass function to
    the analytic Sheth-Tormen 
    mass function.}
  \label{fig:baryon_hmf}
\end{figure}
Our simulation contains a large enough field volume to allow us to
study the abundance of the low-mass halos important for
lensing. Fig.~\ref{fig:baryon_hmf} shows the measured halo mass
function in both the hydrodynamical and DMO versions of our simulation
at redshift $z=0$. We find that the mass functions in both versions of
the simulation are well fit by a power law, of the form, 
\begin{equation}
    \frac{\mathrm{d}n}{\mathrm{dlog_{10}}M/M_\odot} = b (M/M_\odot)^{-a} \;,
\end{equation}
in the range ($3\times 10^6$ -- $3\times 10^{11}$)~\msun{}. The best fit
parameters are listed in Table~\ref{tab:mf_fits}. We find no
significant difference between the slope of the halo mass functions in
the hydrodynamical and DMO versions of our simulation. Across all halo
masses considered, the amplitude of the DMO mass function is greater
than the amplitude of the hydrodynamical mass function by around
25\%.
Given the mass function is a power-law with a slope of approximately -1, this difference is equivalent to all halos in the DMO simulation
having their mass reduced by approximately 25\%, consistent with the reduction in halo mass (at low halo masses) going from DMO to \eagle{} shown in \citet{2015MNRAS.451.1247S}.\footnote{Different hydrodynamical simulations disagree slightly on the effects of baryons on the halo mass function. For example, GIMIC was similar to \eagle{}, with a roughly 30\% reduction in the mass of $\sim 10^{10} \, M_\odot$ haloes \citep{sawala2013}, while \citet{2018MNRAS.481.1950L} showed that the IllustrisTNG simulations show only a 20\% reduction in mass for similar-mass haloes.}

The reduction in halo mass is caused by two
processes operating at early times. Firstly, after the primordial gas
is reionized, photo-heating evaporates gas from small mass halos or
prevents it from cooling into them. Secondly, in halos where gas does
cool and make stars supernovae expel the remaining gas \citep[][and
references therein]{benson2002,benitez-llambay2020}. Of course, these
processes are not modelled in DMO simulations and DMO halos become
around 15\% more massive (the value of
$\Omega_\mathrm{b}/\Omega_\mathrm{m}$) than an otherwise equivalent
halo in a hydrodynamical simulation. The loss of mass from these
processes reduces the rate at which halos grow in the hydrodynamical
simulation and the 15\% difference at the redshift of reionisation
increases to the 25\% mass difference in halo mass at the present day
\citep{sawala2016}.

The measured slope of the halo mass function is shallower than the
slope of the Sheth-Tormen mass function --- 0.90 in the simulation and
0.92 in the Sheth-Tormen model. We can see in the lower panel of 
Fig.~\ref{fig:baryon_hmf} that the Sheth-Tormen model overpredicts the
abundance of halos less massive than $10^{10}$~\msun{} in our high-resolution volume. 
While the difference in abundance between the Sheth-Tormen prediction and the DMO simulation could be affected by the special nature of the volume we have simulated, the difference in slope seems to be robust, as is the difference between the DMO and hydrodynamical simulation.
We
therefore conclude that previous studies which used the Sheth-Tormen
model, e.g. \citet{li2017}, may have overpredicted the expected lensing
signal originating from halos in the $10^7$--$10^8$~\msun{} range by
around 20--30\%. Whilst we are unable to check whether the same
overprediction applies to the calculation of the lensing signal in a
WDM cosmology, this difference in the expected abundance of halos in a
CDM universe is important from an observational standpoint.
\begin{table}
\centering
\begin{tabular}{@{}lcc@{}}
\toprule
      & $a$              & $b$ [Mpc$^{-3}$] \\ \midrule
Hydro & $0.897\pm0.005$ & $2.2\pm0.2\times10^8$    \\
DMO   & $0.898\pm0.009$ & $2.8\pm0.5\times10^8$    \\ \bottomrule
\end{tabular}%
\caption{Slope and amplitude of power-law fits to the halo mass function
        in the high-resolution region of our simulation at redshift $z=0$.}
\label{tab:mf_fits}
\end{table}

\subsection{The effect of environment on the halo mass function}
We also study the effect of environment on the abundance and
properties of field halos. We use the \nexus{} code \citep{cautun2013}
to classify halo environments. \nexus{} divides space into a cubic
grid, and classifies each cell as belonging to either a void, a sheet,
a filament, or a node. The method is scale-free, analysing the density
field smoothed on a number of different scales in order to detect
structure of all sizes.

\begin{figure*}
  \includegraphics[width=\linewidth]{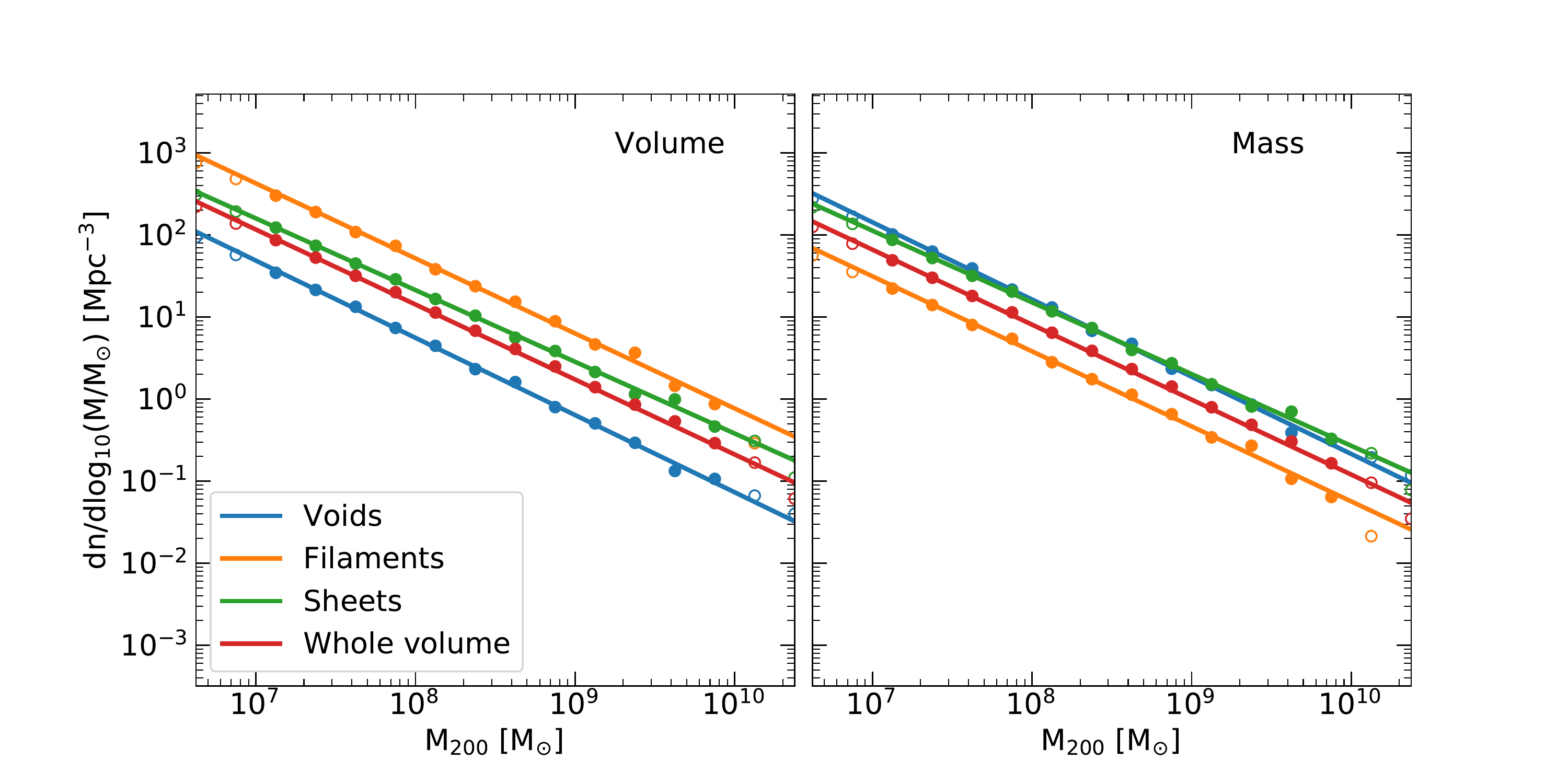}
  \caption{The differential halo mass function for halos in voids
    (blue), filaments (orange), sheets (green), and the entire volume
    (red) in the high-resolution region of the hydrodynamical version
    of our simulation at redshift $z=0$. The environment of a halo is
    determined using the \nexus{} algorithm
    \citep{cautun2013}. Circles show the measured mass function,
    whilst lines show power-law fits. In the left-hand panel the
    amplitude of the mass function is normalised to the physical
    volume of each environment type. Empty circles show points not
    used when calculating power-law fits. In the right-hand panel the
    amplitude is normalised to the Lagrangian volume of each
    environment type, i.e the mass contained in each environment
    type.}
  \label{fig:environment_hmf}
\end{figure*}

The mass function of halos in voids, sheets and filaments is shown in
the left-hand panel of Fig.~\ref{fig:environment_hmf}. The slope of
the mass function does not depend strongly on halo environment, but
the amplitude of the mass function in different environments is
strongly correlated with the average density of those environments;
the amplitude of the halo mass function in filaments is an order of
magnitude greater than in voids. It is natural to wonder whether the
difference in amplitude results solely from the difference in the
density of matter in each region. To account for the differing
densities in each environment type, we also calculate the halo mass
function per unit Lagrangian volume\footnote{The Lagrangian volume
  represents the comoving volume which would have been occupied by a
  region at the Big Bang, and can be calculated by dividing the total
  mass of matter in a region by the mean matter density of the
  Universe.}. The results are shown in the right-hand panel of
Fig.~\ref{fig:environment_hmf}. We see here that relative to the
density of matter in each region, halos in the mass range considered
here are less abundant in filaments than in voids. The halo masses we
consider all lie comfortably below the characteristic clustering mass
scale, $M^{\star}(z)$, which at redshift $z=0$ is around
$6\times10^{12}$~\msun{} \citep{white1993, schneider_m_2012}. The abundance of
halos of a fixed mass below $M^{\star}(z)$ eventually decreases in
time, as these smaller halos merge and accrete material to become
larger halos. The higher density filament regions are effectively in a
more advanced state of cosmic evolution relative to the lower
density void regions, so the abundance of halos less massive than
$M^{\star}(z)$ ends up lower in the filaments.

\subsection{The effect of environment on the internal structure of halos}
\begin{figure}
  \includegraphics[width=\linewidth]{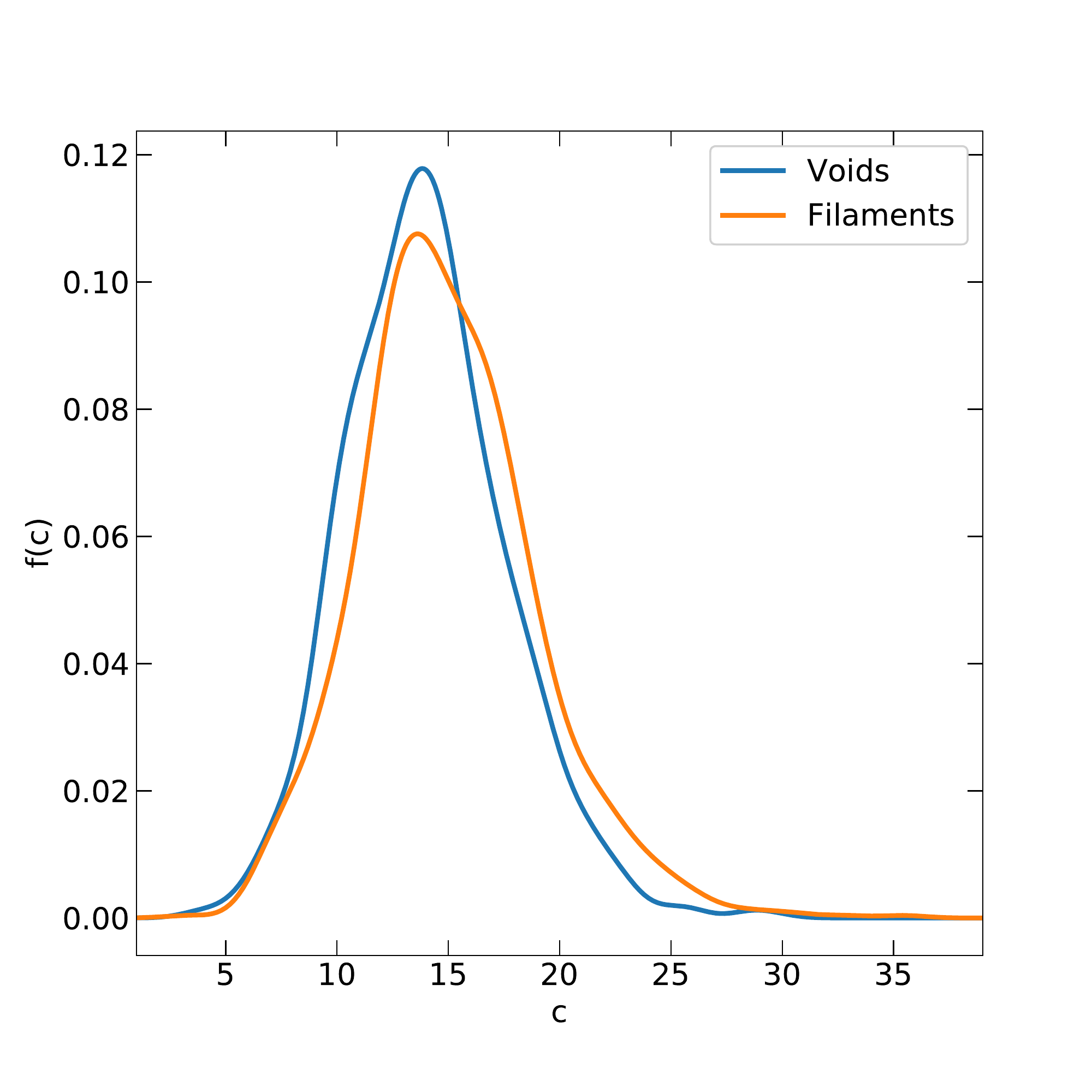}
    \caption{The distribution of concentration for halos in filaments and voids
    in the hydrodynamical version of our simulation at redshift
    $z=0$. Halos are selected to have masses between
    $10^{7.5}$--$10^{9.5}$~\msun{}. All halos in our sample satisfy
    the three relaxation criteria of \citet{neto2007}, and the
    concentrations are calculated by fitting NFW profiles.} 
  \label{fig:conc_dist}
\end{figure}

We also consider the relationship between halo environment and the
internal structure of the halo. Specifically, we compare the
concentrations of halos in voids and filaments, for halo masses
between $10^{7.5}$--$10^{9.5}$~\msun{}. If the halo has an NFW density
profile \citep{navarro1996,navarro1997}, with scale radius,
$r_\mathrm{s}$, the concentration, $c$, is given by
$r_{200}/r_\mathrm{s}$.
We only consider halos which satisfy the three relaxation criteria of
\citet{neto2007}, and where $r_\mathrm{s}$ is greater than the
convergence radius of the halo, as defined using the criterion of
\citet{power2003}. The distribution of concentrations for halos in the
mass range $10^{7.5}$--$10^{9.5}$~\msun{} at redshift $z=0$ is shown
in Fig.~\ref{fig:conc_dist}. 
Whilst the width and skew of the distribution is similar in both
filaments and voids, we see that halos in filaments tend to have
slightly higher concentrations, and halos with a concentration greater
than 25 reside exclusively in filaments. The concentration of a halo
reflects the density of the universe at its formation time
\cite{navarro1997}. For a fixed mass, halos tend to form earlier in
filaments than voids \citep{hahn2007}, when the universe was
denser. This explains the higher average concentration observed for
halos in filaments.

\section{The subhalo population}
\label{section:subhalos}

The small dark matter particle mass of our simulation allows us to
study the abundance and properties of subhalos as small as
$10^7$~\msun{}. This is the first time that such small substructures
have been studied in a hydrodynamic simulation of a $10^{13}$~\msun{}
halo.  In this section we focus on how the inclusion of baryons in the
simulation changes the abundance and properties of this subhalo
population.

For low-redshift halos of mass ${\sim}10^{13}$~\msun{}, a
significant fraction of the distortions to strong lensing arcs is due
to substructure within the lensing halo. For example, for a typical
SLACS lens (at $z=0.2$, with a source at $z=1$), CDM substructure
produces around 30\% of the lensing distortions, whilst in WDM the
contribution of substructures is comparable to that from field halos
along the line-of-sight \citep{li2017,despali2018}. 

\subsection{The subhalo mass function}
\begin{figure*}
  \includegraphics[width=\linewidth]{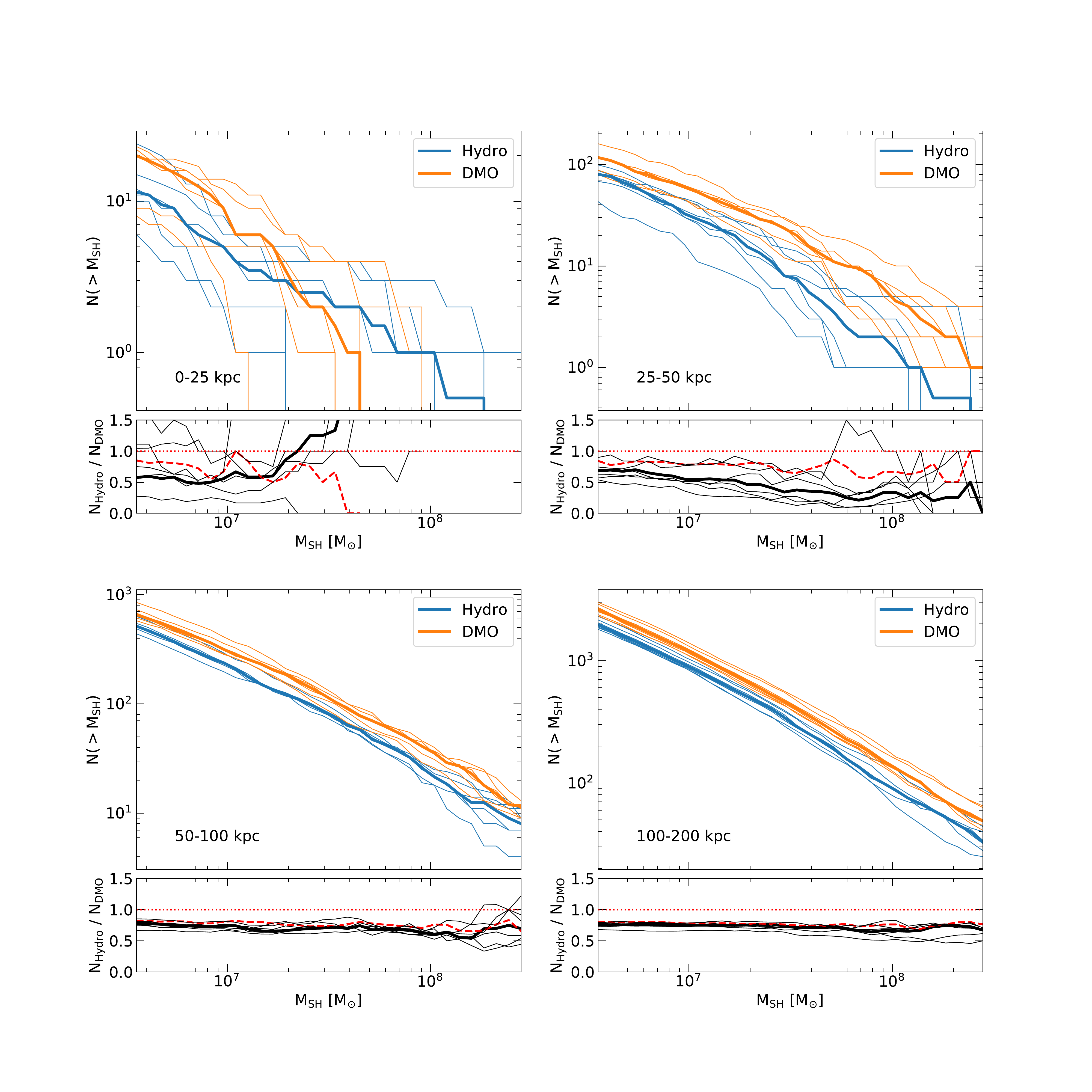}
  \caption{Large panels: cumulative subhalo mass functions in
    concentric spherical shells centred on the potential minimum of
    the central halo. Thin lines show the abundance of subhalos at six
    individual snapshots, approximately evenly spaced in time between redshift $z=0.5$ and
    the present day.  Thick lines show the abundance of subhaloes
    averaged over these six snapshots.  Small panels: the ratio of the
    cumulative subhalo mass functions in the hydrodynamical and DMO
    versions of the simulation at each snapshot (thin black
    lines). The thick black lines show the average reduction in
    subhalo abundance as a function of mass over a 5~Gyr period. The
    dashed red lines show the reduction in subhalo abundance
    when the masses of the objects in the DMO simulation are
    multiplied by 0.75 to approximate the reduced-growth effect
    described in \S\ref{section:baryon_mf}.}
  \label{fig:subhalo_mf}
\end{figure*}

Fig.~\ref{fig:subhalo_mf} shows the cumulative subhalo mass function
in four concentric spherical shells centred on the potential minimum
of the halo. We see that the inclusion of baryons in the simulation
leads to a reduction in subhalo abundance as a function of subhalo
mass. As discussed in \S\ref{section:baryon_mf}, halos in cosmological
hydrodynamical simulations are systematically less massive than their
DMO counterparts because the loss of baryons at early time reduces
their subsequent growth rate. To distinguish this ``reduced-growth''
effect from environmental effects, such as tidal stripping and
disruption, we apply a correction to the subhalo abundance in the DMO
simulation by reducing the masses by 25\%, which is the typical size
of the reduced-growth effect. The corresponding reduction in subhalo
abundance is shown by the red dotted line in each panel.

In the innermost radial bin, the total number of subhalos in the mass
range $(3\times 10^{6}-3\times 10^{7})$~\msun{} is reduced by around
50\%, although there is considerable scatter in the different
snapshots. Approximately half the measured reduction is due to
dynamical processes -- tidal stripping and destruction -- and half to
the reduced-growth effect. The average reduction in subhalo abundance
in this region is comparable to that in Milky Way-mass halos found in
the \apostle{} simulations\footnote{In general, the galaxy mass--halo
  mass relation peaks at a mass of $10^{12}$~\msun{}; however the
  galaxies in the \apostle{} simulations are unusually small for their
  halo size.}, which also used the \eagle{} model
\citep{richings2018}. This is not surprising as the ratio of galaxy to
halo mass is similar in all these simulations.

There is a clear radial trend in the reduction of subhalo abundance in
the hydrodynamical simulation.  The effect of the central galaxy on
the subhalo population is negligible at distances greater than 100~kpc
(which is also the case in the \apostle{} simulations). Here, the
reduction is essentially independent of subhalo mass and is explained
entirely by the reduced-growth effect in hydrodynamical simulation. In
the inner shells, where the effect of the central galaxy is important,
there seems to be some dependence of the reduction on subhalo mass but
the numbers are too small to reach a firm conclusion.

\subsection{Subhalo concentrations}

Since the size of a subhalo is not well defined, it is better to
characterise their concentrations in terms of their mean overdensity,
\dv{}, within the radius, \rmax{}, at which the circular velocity
peaks, in units of the critical density,
\begin{equation}
    \delta_V = 2\left(\frac{V_{\mathrm{max}}}{H_0 r_{\mathrm{max}}}\right)^2 \;,
\end{equation}
where \vmax{} is the maximum circular velocity of the
halo\footnote{$V_{\mathrm{max}}=\mathrm{max}\left(\sqrt{\frac{GM(<r)}{r}}\right)$}
\cite{springel2008}. 
For an NFW halo, the concentration, $c$, is related to \dv{} by
\begin{equation}
    \delta_V = 7.213\left(\frac{200}{3}\right)\frac{c^3}{\ln(1+c) - c/(1+c)} \;.
\end{equation}
Whilst this equation cannot be inverted analytically, we find that an
approximate relation that holds well for concentrations between 5 and 50 is
\begin{equation}
    c = 0.3\delta_V^{0.4} \;.
\end{equation}
\begin{figure}
  \includegraphics[width=\linewidth]{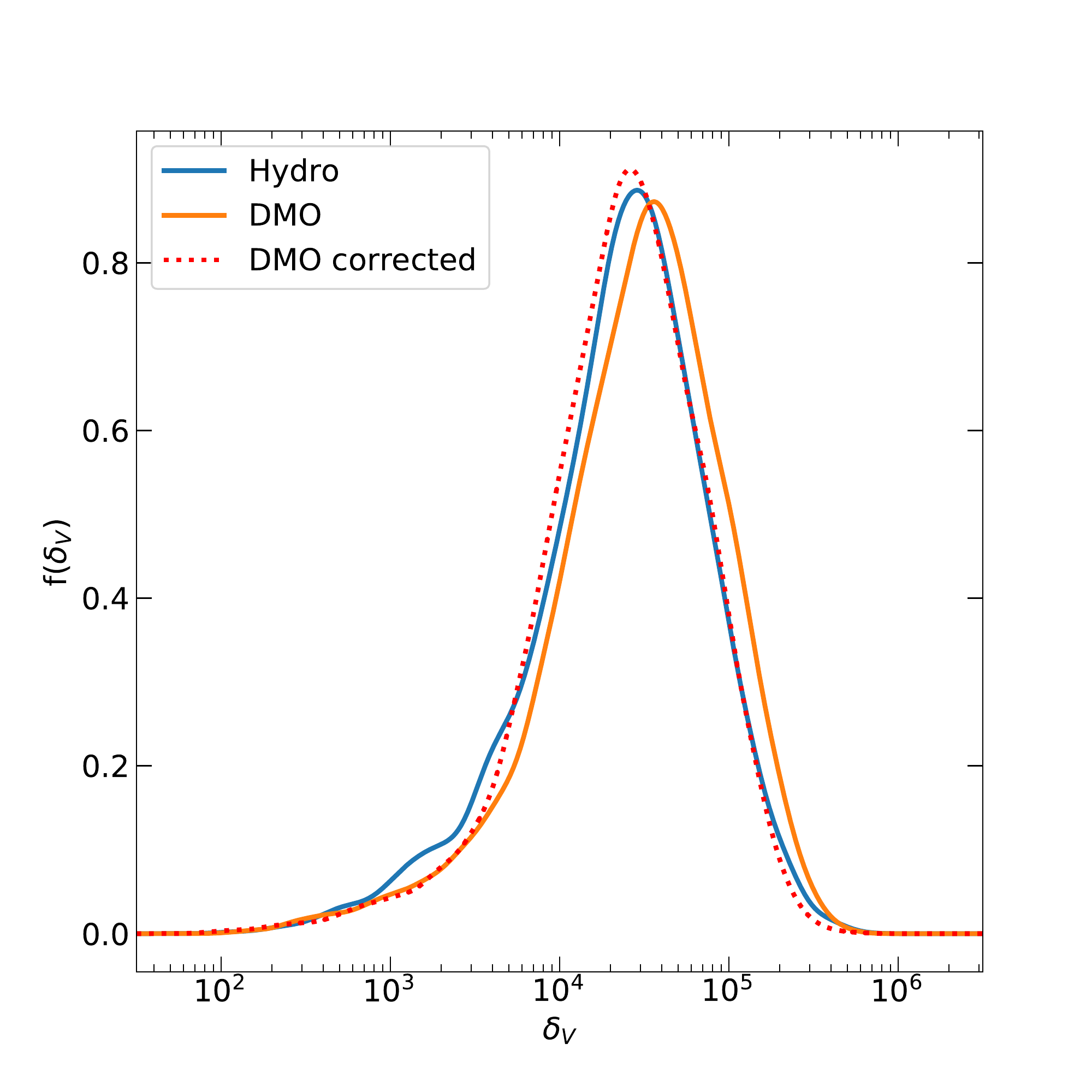}
  \caption{The distribution of subhalo characteristic overdensity,
    \dv{} (which we use to characterize subhalo concentration) in
    the hydrodynamical and DMO versions of our simulation at redshift
    $z=0$. Subhalos are selected to have maximum circular velocities
    between 3 and 20 km/s. The dotted red line corresponds to the case
    when the DMO \vmax{} values are multiplied by 0.85 to account for
    the systematic mass difference between hydrodynamical and DMO
    halos due to the reduced-growth effect discussed in
    \S\ref{section:baryon_mf}.}
  \label{fig:subhalo_conc_dist}
\end{figure}

The distribution of \dv{} for subhalos with \vmax{} between 3 and 20
km/s lying within 500~kpc of the centre of the main halo at $z=0$ is
shown in Fig.~\ref{fig:subhalo_conc_dist}. We only consider
well-resolved subhalos by requiring that \rmax{} be greater than the
gravitational softening length, 0.5 kpc.
Subhalos in the hydrodynamical version of our simulation are
systematically less concentrated than subhalos in the DMO version,
although the difference is small. The peak of the DMO distribution
occurs at a value of \dv{} which is 23\% higher than in the
hydrodynamical simulation. The difference in \dv{} is equivalent to a
difference of approximately 8\% in concentration for NFW halos.
Fig.~\ref{fig:subhalo_conc_dist} also shows the distribution of \dv{}
for subhalos in the DMO simulation when the values of \vmax{} are
reduced by 15\% to mimic the reduced-growth effect discussed in
\S\ref{section:baryon_mf}, as found by \citet{sawala2016} for field
halos. This slight shift in \vmax{} largely explains the difference
between the hydrodynamics and DMO distributions. We conclude that the
inclusion of baryons in the simulations does not have a significant
impact on the concentration of subhalos in the mass range considered,
beyond a small shift.

\subsection{Projection effects}
The projected mass distribution is responsible for gravitational
lensing and, since the spatial distribution of mass around a large
halo is strongly anisotropic, the observed lensing effect will depend
on the direction along which the lens is observed.  The central halo
in our simulation sits at the intersection of three filaments (see
Fig.~\ref{fig:simpic}). The number density of substructures along
these filaments is greater than the average around the halo, so a lens
observed along a a filament will be affected by substructure much more
strongly than a lens observed along an average direction.

\begin{figure*}
  \includegraphics[width=\linewidth]{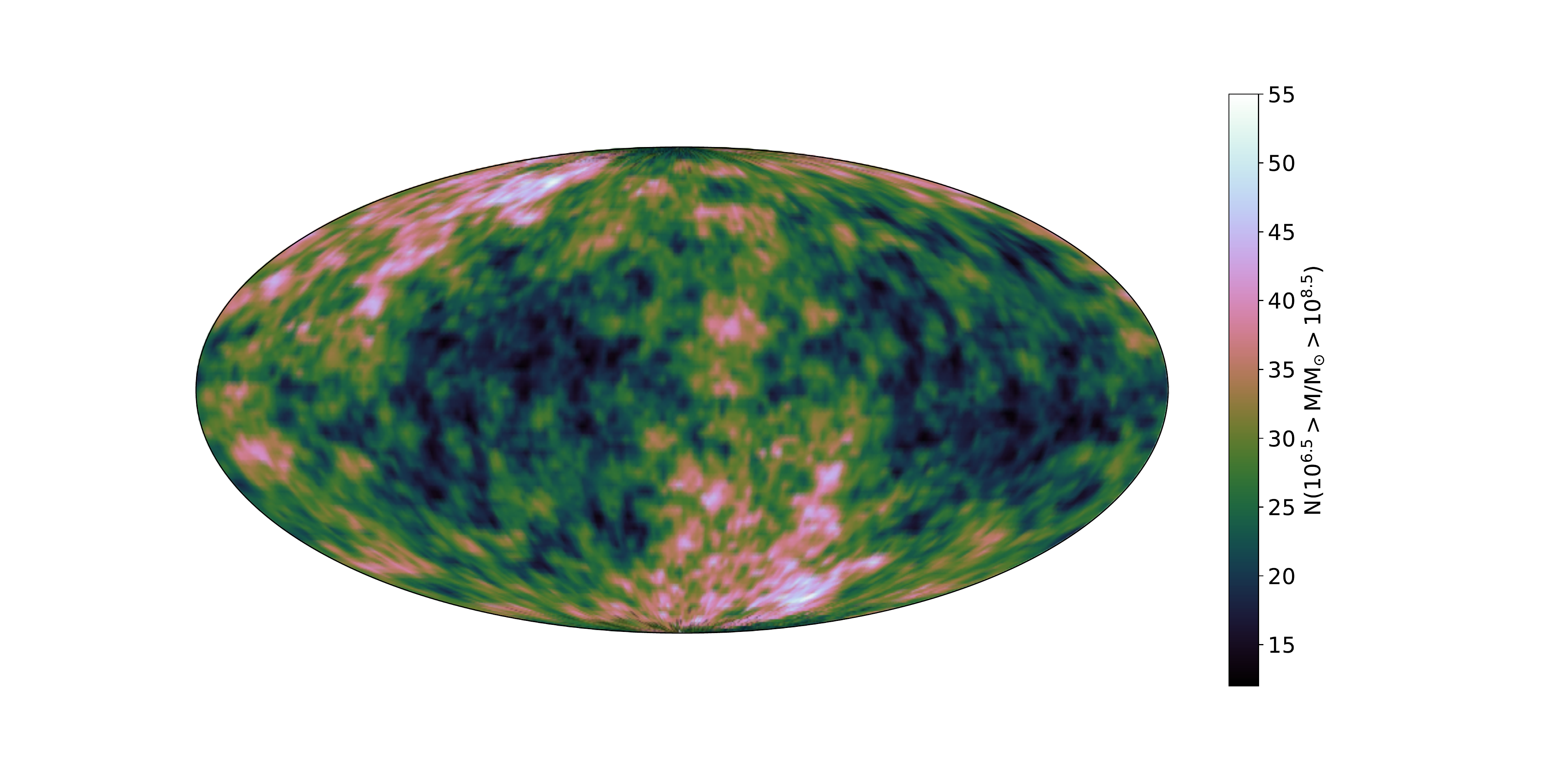}
  \caption{The number of halos and subhalos of mass in the range
    $(10^{6.5}$--$10^{8.5})$~\msun{} along lines-of-sight to the main
    cluster in the hydrodynamical simulation at redshift $z=0.1$.
    Each line-of-sight is a cylinder of 10~Mpc length and 10~kpc
    radius.  The map is an equal-area Mollweide projection, smoothed
    on a scale of one degree, made from
    $10^6$ lines-of-sight spread almost uniformly across the surface
    of a sphere of radius 5~Mpc centred on the main halo.}
  \label{fig:nsub_proj}
\end{figure*}

A visual representation of the dependence of the observed abundance of
substructure on viewing angle is presented in
Fig.~\ref{fig:nsub_proj}. To construct this image we distributed
$10^6$ lines-of-sight uniformly on the surface of a
sphere\footnote{Technically, an exactly uniform spacing of points on
  the surface of a sphere is impossible for all but a set of special
  numbers of points\citep{saff1997}. Here we used the python package
  \textsc{Seagen} \citep{kegerreis2019} to distribute points on the
  surface of a sphere such that the density of points over the sphere
  is very close to uniform, including at the poles.}  centred on the
potential minimum of the main halo. Along each line-of-sight, we
calculate the number of halos and subhalos with a \subfind{}
mass\footnote{That is the mass found by the \subfind{} algorithm
  \citep{springel2001} which, for subhalos, corresponds closely to the
  mass enclosed by the tidal radius \citep{springel2008}} between
$10^{6.5}$--10$^{8.5}$~\msun{}, in a cylinder of radius 10~kpc and
length 10~Mpc centred on the main halo. This includes the
subhalos of the main halo, and also other halos and their subhalos
which fall along the line-of-sight. The map of the number of objects
along each line-of-sight in Fig.~\ref{fig:nsub_proj} is smoothed on a
scale of one degree and is for the cluster at redshift $z=0.1$ since
this is typical of low-redshift lenses \citep[e.g.][]{bolton2006} and
is the value used in the analysis of \citet{li2017}.

It is clear that the number of objects varies strongly with viewing
angle. Highly populated viewing angles are closely aligned
with filaments and often contain 2--3 times as many objects as
viewing angles that do not overlap a filament. 
The dominant contribution to the signal originates from subhalos, not
from nearby field halos although the distinction between halos and
subhalos is ambiguous as the shape of the halo, and thus the number of
subhalos along a particular line-of-sight, is strongly correlated with
the direction of the filaments. From an observational perspective, the
distinction is artificial.
\begin{figure}
  \includegraphics[width=\linewidth]{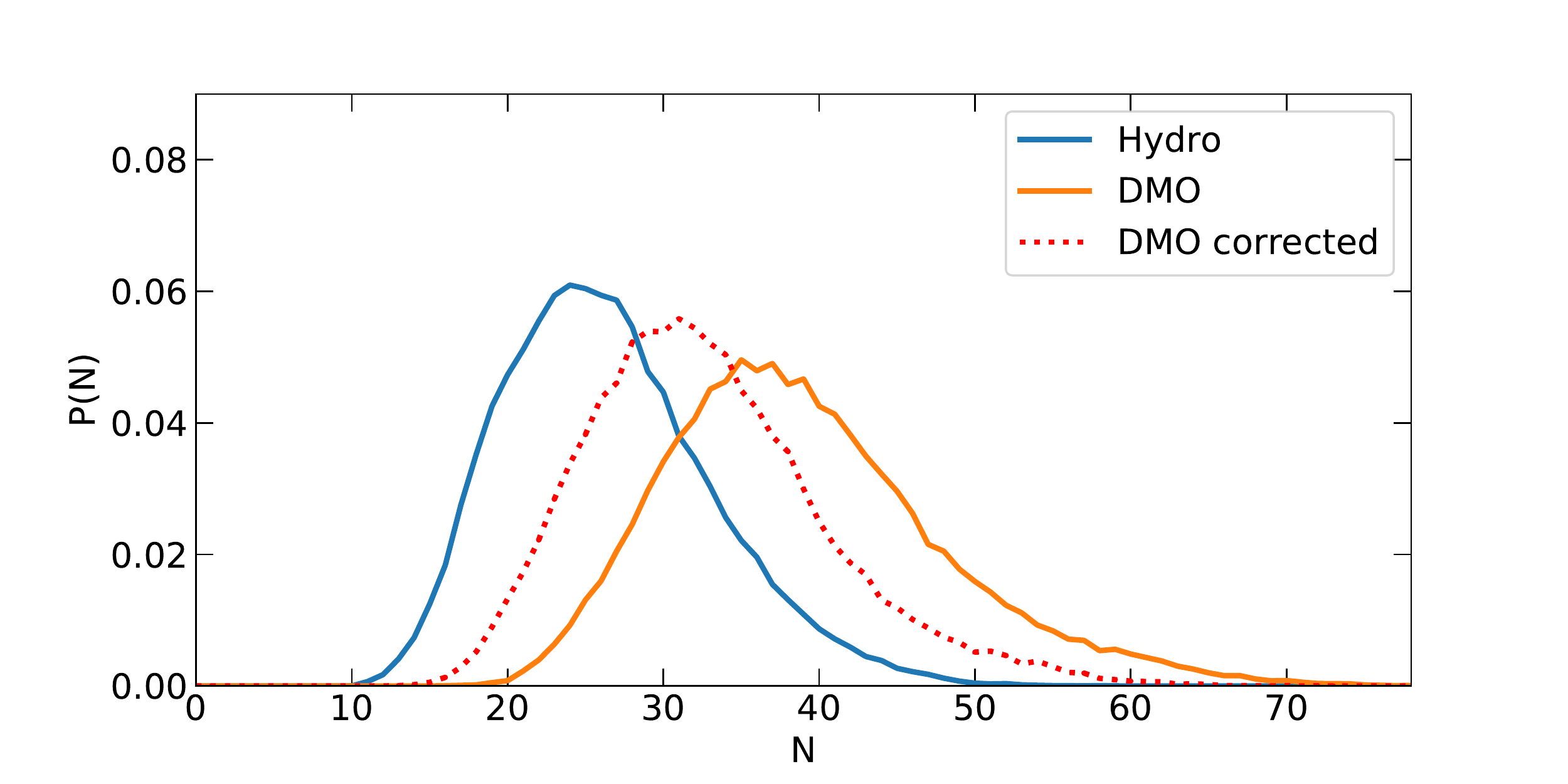}
  \caption{The distribution of the number of halos and subhalos of
    mass between $10^{6.5}$--$10^{8.5}$~\msun{} along lines-of-sight
    projected through the centre of the main halo at redshift
    $z=0.1$. Each projection is of a cylinder of 10~Mpc length and
    10~kpc radius. The dotted red line shows the distribution in
    the case where the masses of all objects in the DMO version of our
    simulation are multiplied by 0.75 to account for the effect
    discussed in \S\ref{section:baryon_mf}.}
  \label{fig:nsub_dist}
\end{figure}

We compare the distribution of the number of objects along different
lines-of-sight in the hydrodynamical and DMO versions of our
simulation in Fig.~\ref{fig:nsub_dist}. The median number of objects
along a line-of-sight, and the interquartile ranges, are listed in
Table~\ref{tab:proj_counts}. The number of objects along a given
line-of-sight in the hydrodynamical simulation is around 30\% smaller
on average. This is a combination of the reduced-growth effect 
together with the destruction and tidal stripping of subhalos in the
hydrodynamical simulation. Comparison of the abundance in the
hydrodynamical simulation to that in the DMO simulation with the
masses of objects reduced by 25\% (dotted red line) shows that the
reduced-growth effect accounts for approximately half of the measured
difference between the hydrodynamical and DMO simulations. In the
hydrodynamical simulation, the median number of objects along the line
of sight is 26, but there are lines-of-sight that intercept more than
twice this number.
\begin{table}
\centering
\begin{tabular}{@{}lc@{}}
\toprule
Simulation      & N              \\ \midrule
Hydro           & $26^{+5}_{-4}$ \\
DMO             & $38^{+5}_{-5}$ \\
DMO - corrected & $32^{+5}_{-5}$ \\ \bottomrule
\end{tabular}
\caption{The median number of objects in the mass range
  $(10^{6.5}$--$10^{8.5})$~\msun{} along a 10~Mpc long
  cylindrical line-of-sight of radius 10~kpc centred on the
  potential minimum of the main halo. Subscripts
  and superscripts give the interquartile range. The numbers quoted 
  includes subhalos of the main halo as well as field halos.}
\label{tab:proj_counts}
\end{table}

\section{Conclusions}
\label{section:conclusions}

We have developed a new technique to generate initial conditions for
cosmological smooth particle hydrodynamics simulations in which the
number of dark matter particles can be much larger than the number of
gas particles. Our main motivation is to simulate a massive elliptical galaxy
with realistic galaxy formation astrophysics -- which requires good gas
resolution -- while, at the same time, resolving the
$\sim 10^6$~\msun{} haloes and subhaloes relevant to strong
gravitational lensing tests of the identity of the dark matter --
which requires very high dark matter resolution. An added benefit of
our new technique is that it avoids the 2-body scattering processes
inherent in the traditional cosmolgical SPH setup in which the dark
matter and the gas are followed with the same number of particles 
which, consequently have very different masses \citep{ludlow2019}.

We have simulated a $10^{13}$\msun{} galaxy cluster and its
surrounding large-scale environment, a volume of over 500 Mpc$^3$,
using the \eagle{} \textsc{Reference} model of galaxy formation. Our
conclusions may be summarized as follows:

\noindent $\bullet$ The field halo mass function in the mass range
($5\times 10^6 - 3 \times 10^{11}$)~\msun{} closely follows a power
law of slope -0.9 in both the DMO and hydrodynamic simulations (see
Table~\ref{tab:mf_fits}). However, the amplitude of the halo mass
function in the hydrodynamics case is about 25\% lower than in the DMO
case (Fig.~\ref{fig:baryon_hmf}). The difference originates at early
times when halos in the hydrodynamics simulation lose gas, either as a
result of reionization or of supernovae feedback and, as a result,
experience less growth than their DMO counterparts, as first discussed
by \cite{sawala2016}.

\noindent $\bullet$ The halo mass functions are not well described by
the commonly used Sheth-Tormen formula, which is based on a fit to DMO
simulations and has a steeper slope than we measure. As a result,
previous lensing studies using the Sheth-Tormen model have
overpredicted the expected lensing signal originating from halos in
the ($10^7$--$10^8$)~\msun{} range by around 20--30\%.

\noindent $\bullet$ The abundance of field halos depends sensitively on
environment. In our hydrodynamical simulation we find that the number
of halos per unit mass in the range of halo masses considered here is
largest in the sheets and voids of the cosmic web, where it exceeds the number
per unit mass in filaments by a factor of four to five (although
the volume-weighted number is largest in filaments;
Fig.~\ref{fig:environment_hmf}).

\noindent $\bullet$ The mass function of {\em subhalos} in the cluster
also has lower amplitude in the hydrodynamical simulation than in the
DMO simulation (Fig.~\ref{fig:subhalo_mf}). In addition to the same
reduced growth experienced by field halos, the subhalo abundance is
further reduced in the hydrodynamical simulation by the enhanced
destruction of subhalos caused by the stronger tidal interactions in
the presence of a massive galaxy at the centre of the cluster. The
extent of this destruction depends sensitively on radius. For example,
within 50~kpc in projection, the number of substructures in the
($10^{6.5}$--$10^{8.5}$)~\msun{} mass range in the hydrodynamics
simulation is only about half the number in the DMO simulation
(with considerable halo-to-halo scatter). Approximately 50\% of this
difference is accounted for by the reduced growth effect in the
hydrodynamical simulation and the remaining 50\%  by tidal
disruption. Beyond 100~kpc from the centre, the effect of the central
galaxy is small and the reduction is due almost entirely to the
reduced-growth effect.

\noindent $\bullet$ Subhalos in the hydrodynamical simulation are less
concentrated than their DMO counterparts but the difference is only
about 10\%. It arises from the reduced-growth effect which
effectively shifts the formation time of halos in the hydrodynamical
simulation to slightly later times.

\noindent $\bullet$ The matter distribution around the cluster is
highly anisotropic and, as a result, the projected number of halos and
subhalos -- the quantity of interest in strong gravitational lensing
studies -- is also highly anisotropic. For example, the projected
number of objects in the mass range ($10^{6.6}-10^{8.5}$)~\msun{} along
a cylinder of radius 10~kpc and length 10~Mpc centred on cluster can
be 2-3 times larger if aligned with a filament than if not.

The analysis of the perturbations on strong gravitational lenses
offers a real prospect of testing the $\Lambda$CDM model in the regime
of small-mass halos where it makes robust predictions that distinguish
it from viable alternatives such as WDM \citep{li2017}. The prime
targets for this kind of lensing studies are $10^{13}$~\msun{}
halos like the one we have simulated here. Understanding the
abundance, structure and distribution of subhalos in these halos, 
and of field halos around them, is an important prerequisite for the
successful application of lensing techniques to the problem of the
identity of the dark matter. 

\section*{Acknowledgements}

We thank an anonymous referee for a positive and constructive review.
We acknowledge support from the European Research Council
through ERC Advanced Investigator grant, DMIDAS [GA 786910] to
CSF. This work was also supported by STFC Consolidated Grants for
Astronomy at Durham ST/P000541/1 and ST/T000244/1. AR is supported by the European Research Council's Horizon2020 project `EWC’ (award AMD-776247-6). It used the
DiRAC Data Centric system at Durham University, operated by the
Institute for Computational Cosmology on behalf of the STFC DiRAC HPC
Facility (\url{www.dirac.ac.uk}). This equipment was funded by BIS National
E-infrastructure capital grants ST/P002293/1, ST/R002371/1 and
ST/S002502/1, Durham University and STFC operations grant
ST/R000832/1. DiRAC is part of the National e-Infrastructure. 

\section*{Data availability}

The data underlying this article will be shared on reasonable request to the corresponding author.

\bibliographystyle{mnras}
\bibliography{bibliography}

\end{document}